\documentclass[iop,apjl]{emulateapj}
\newcommand{\V}[1]{\mathbf{#1}} %Bold for Vectors
\newcommand\Alfven{Alfv\'en } %Proper Names
\newcommand\Alfvenic{Alfv\'enic } %Proper Names
\usepackage{amssymb}
\usepackage{amsmath}
\usepackage{hyperref}
\usepackage{natbib}
\usepackage{color}
\usepackage{graphicx}% Include figure files
\begin{document}
\title{The Violation of the Taylor Hypothesis in Measurements of Solar Wind Turbulence}
%\author{K.~G. Klein}
%\affil{
%Space Science Center, University of New Hampshire, Durham,
%NH 03824, USA
%}
%\email{kristopher.klein@unh.edu}
%\author{G.~G. Howes}
%\affiliation{
%Department of Physics and Astronomy, University of Iowa, Iowa City,
%IA 52242, USA.
%}
%\author{J.~M. TenBarge}
%\affiliation{
%IREAP, University of Maryland, College Park, MD 20742, USA.
%}

\author{K.~G. Klein$^1$,
G.~G. Howes$^2$,
J.~M. TenBarge$^{3}$}

\affiliation{$^1$Space Science Center, University of New Hampshire, 
Durham, NH 03824, USA\\
$^2$Department of Physics and Astronomy, University of Iowa, 
Iowa City, IA 52242, USA.\\
$^3$IREAP, University of Maryland, College Park, MD 20742, USA.}

\begin{abstract}
Motivated by the upcoming \emph{Solar Orbiter} and \emph{Solar Probe
  Plus} missions, qualitative and quantitative predictions are made
for the effects of the violation of the Taylor hypothesis on the
magnetic energy frequency spectrum measured in the near-Sun
environment.  The synthetic spacecraft data method is used to predict
observational signatures of the violation for critically balanced
Alfv\'enic turbulence or parallel fast/whistler turbulence. The
violation of the Taylor hypothesis can occur in the slow flow regime,
leading to a shift of the entire spectrum to higher frequencies, or in
the dispersive regime, in which the dissipation range spectrum
flattens at high frequencies. It is found that Alfv\'enic turbulence
will not significantly violate the Taylor hypothesis, but whistler
turbulence will.  The flattening of the frequency spectrum is
therefore a key observational signature for fast/whistler turbulence.
\end{abstract}


\maketitle
%-=-=-=-=-=-=-=-=-=-=-=-=-=-=-=-=-=-
%-=-=-=-=-=-=-=-=-=-=-=-=-=-=-=-=-=-
\section{Introduction} 

The turbulent cascade of energy from large to
small scales influences plasma evolution and heating in many
astrophysical environments, from galaxy clusters and accretion disks
to the solar corona and solar wind.  Extensive \emph{in situ}
observations of the near-Earth solar wind provide invaluable
opportunities to test theories of turbulent transport, dissipation,
and heating.  Upcoming missions, including \emph{Solar Orbiter} and
\emph{Solar Probe Plus}, will make the first \emph{in situ}
measurements of turbulence in the near-Sun environment, providing crucial data to
identify the mechanisms governing coronal heating.

The interpretation of \emph{in situ} measurements of plasma turbulence
is complicated by the fact that the turbulence is measured in a frame
of reference (the spacecraft frame) that is in relative motion with
respect to the frame of reference of the solar wind plasma (the plasma
frame).  For a spatial Fourier mode with wavevector $\V{k}$, the
transformation from the frequency $\omega$ in the plasma frame to the
observed frequency $\omega_{sc}$ in the spacecraft frame yields the
relation
\begin{equation}
\omega_{sc}= \omega + \V{k}\cdot \V{v}_{sw},
\label{eqn:map}
\end{equation}
\citep{Taylor:1938};
a derivation can be found in a companion work, \cite{Howes:2014a},
heretofore referred to as Paper I.
 For a turbulent
 distribution of modes in wavevector space, the plasma-frame frequency
 term $\omega$ and spatial advection term $\V{k}\cdot \V{v}_{sw}$,
 both contributing to the spacecraft-frame frequency, cannot be uniquely
 separated using single-point spacecraft measurements.

The typically super-\Alfvenic velocity of the solar wind near Earth,
$v_{sw}\gg v_A$, motivates the use of the Taylor hypothesis
\citep{Taylor:1938,Fredricks:1976}, assuming that $|\omega| \ll
|\V{k}\cdot \V{v}_{sw}|$, thereby relating the spacecraft-frame
frequency directly to the wavenumber of spatial fluctuations,
$\omega_{sc}\simeq \V{k}\cdot \V{v}_{sw}$.  When the plasma-frame
frequency is non-negligible, $|\omega| \gtrsim |\V{k}\cdot
\V{v}_{sw}|$, the Taylor hypothesis is violated.  
A number of previous studies have addressed this issue in the context of 
different analyses of solar wind measurements
\citep{Fredricks:1976,Matthaeus:1982b,Goldstein:1986,Leamon:1998b,Jian:2009,Perri:2010}. 
In Paper I (Howes et al 2014b), analytic expressions for the validity of 
the Taylor hypothesis are derived for plasma waves relevant to solar wind turbulence.
In this letter, we aim to
determine the qualitative and quantitative effects of the violation of
the Taylor hypothesis on the magnetic energy frequency spectrum in the
solar wind, and in doing so validate expressions derived in Paper I.

\begin{figure}[t]
\includegraphics[width=8.75cm,viewport=7 3 160 140, clip=true]
%{figs/line_final.eps}
{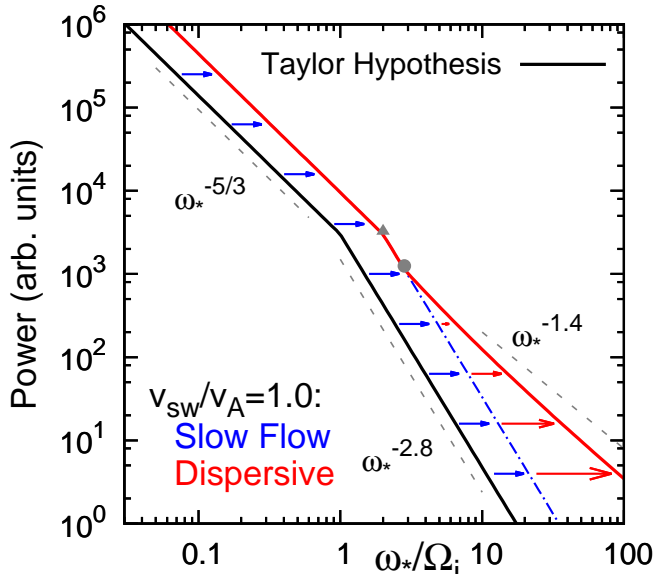}
\caption{ Model of the magnetic energy spectrum vs. normalized
  frequency, $\omega_*/\Omega_i=(\omega_{sc}/\Omega_i)/(v_{sw}/v_A)$,
  for a turbulent distribution of whistler waves with $v_{sw}/v_A=1$,
  illustrating the mapping of the wavenumber spectrum (Taylor
  hypothesis, black) to the measured frequency spectrum. Violation of
  the Taylor hypothesis for slow flow leads to constant shift to
  higher frequency (blue arrows) of the entire spectrum.  At high
  frequency, the dispersive nature of the whistler waves leads to a
  flattening of the frequency spectrum (red arrows).}
\label{fig:line}
\end{figure}

To explore the violation of the Taylor hypothesis for turbulence
measurements in the solar wind requires an estimate of the
plasma-frame frequency of the turbulent fluctuations.  We assume
that the frequency of the turbulent fluctuations is well
characterized by the frequency of the linear waves supported by the
solar wind plasma \citep{TenBarge:2012a}. 
This assumption of linear wave frequencies is one element of a broader 
approach to the modeling of plasma turbulence called the 
\emph{quasilinear premise}; 
a discussion of this approach, including supporting evidence, is presented
in \cite{Klein:2012} and \cite{Howes:2014b}.
Note that the study presented here depends only on the less stringent 
requirement that the linear wave frequency is a good measure of the 
\emph{maximum} frequency of turbulent fluctuations.
In the weakly collisional conditions of the solar wind plasma, we
explore models in which the turbulent fluctuations in the dissipation
range are a broadband spectrum of either kinetic \Alfven waves or
whistler waves, as suggested from a variety of turbulence theories
\citep{Goldreich:1995,Stawicki:2001,Galtier:2006,Boldyrev:2006,Schekochihin:2009},
numerical simulations
\citep{Howes:2008a,Saito:2008,Parashar:2009,Gary:2012,TenBarge:2012a},
and solar wind observations
\citep{Bale:2005,Sahraoui:2010,Salem:2012,Chen:2013a}.

In this Letter, we predict effects of the violation of the Taylor
hypothesis on \emph{Solar Probe Plus} measurements of the magnetic
energy spectrum as a function of the ratio of solar wind velocity to
\Alfven velocity, $\overline{V}\equiv v_{sw}/v_A$.  Both the synthetic
spacecraft data method and a simplified analytical model are used to
predict the mapping of a given wavenumber spectrum of the turbulence
to a measured frequency spectrum.

As illustrated in Fig.~\ref{fig:line}, the Taylor hypothesis is
violated in two regimes: the \emph{slow flow regime} and the
\emph{dispersive regime}.  In the slow flow regime, the solar wind
flow is slow enough that the plasma frame-frequency term is
non-negligible compared to the advection term, $|\omega| \gtrsim
|\V{k}\cdot \V{v}_{sw}|$, leading to a constant shift of the
spacecraft-frame frequency spectrum to higher frequency (blue arrows),
without altering the scaling of the spectrum.  In the dispersive
regime, the plasma-frame frequency increases more rapidly than
linearly with the wavenumber, but the advection term only increases
linearly, so the plasma-frame frequency term will eventually dominate
the spacecraft-frame frequency, leading to a flattening of the
magnetic energy spectrum (red arrows).  This applies to a turbulent
spectrum of kinetic \Alfven waves or whistler waves, but the
anisotropic distribution of turbulent power in wavevector space plays
an important role in distinguishing these two cases.
A complete explanation of Fig.~\ref{fig:line} is deferred to the 
discussion section below.

\section{Synthetic Spacecraft Data Method} To determine the impact of
the plasma-frame frequency on the observed magnetic energy frequency
spectrum, we generate time series using the \emph{synthetic spacecraft
  data method} \citep{Klein:2012}. Adopting the \emph{quasilinear premise},
this method models the turbulence as a
spectrum of randomly-phased, linear kinetic wave modes.  By sampling
along a trajectory through the synthetic plasma volume, we create
single-point time series that may undergo the same analysis as
\emph{in situ} measurements.  First- and second-order correlations in
the turbulence, including energy spectra, can be modeled using these
synthetic time series 
and are found to be in good agreement with solar wind observations
\citep{Howes:2012a,Klein:2012,TenBarge:2012b,Klein:2014a}, but since
the nonlinear interactions responsible for the turbulent energy
transfer between modes are not modeled, such synthetic time series
cannot be used to study third- or higher-order correlations. 

To create the synthetic data, the magnetic field is calculated as a
time series along a defined trajectory $\V{r}(t)=-\V{v}_{sw}t$ in the
plasma volume according to eq.~(3) from Paper I
\begin{equation}
\V{B}(t)  =  \sum_m\sum_{\V{k}} \hat{\V{B}}_m(\V{k}) e^{-i[\V{k}\cdot \V{v}_{sw} + \omega_m(\V{k})]  t},
\label{eqn:Bt}
\end{equation} 
where $\omega_m(\V{k})$ is the linear eigenfrequency for wave mode $m$
with wavevector $\V{k}$, and $\hat{\V{B}}_m(\V{k})$ is the
corresponding complex Fourier coefficient, each of which is multiplied
by a random phase $\exp(i\eta_{\V{k},m})$.

In this Letter, the synthetic data is generated on a cubic
wavevector grid with $512^3$ points, spanning one of two ranges: the
inertial range $0.01 \le |k_j \rho_i| \le 2.56$, or the transition range
$0.1 \le |k_j \rho_i| \le 25.6$, where $\rho_i$ is the ion gyroradius
and the index $j$ signifies $x$, $y$, or $z$.
The frequencies and eigenfunctions for the linear kinetic wave modes
are calculated numerically for each wavevector using the
linear Vlasov-Maxwell dispersion relation
\citep{Quataert:1998,Howes:2006}.  The fully ionized, proton-electron
plasma is assumed to have an isotropic, non-relativistic Maxwellian
velocity distribution with a realistic mass ratio, $m_i/m_e=1836$, and
equal ion and electron temperatures.  We choose the ion
plasma beta $\beta_i= 8 \pi n_iT_i/B^2=1$, so the ion gyroradius and
ion inertial length are equal, $d_i=\rho_i /\sqrt{\beta_i} $ 

We study the violation of the Taylor hypothesis for two turbulence
models: (i) critically balanced \Alfvenic turbulence, or (ii) 
parallel fast/whistler turbulence.  For the \Alfvenic case, only
wavevectors at or below critical balance, 
$k_\parallel  B_0 \lesssim k_\perp \delta B_\perp$,
\citep{Goldreich:1995,Howes:2008b,Howes:2011c,TenBarge:2012a} 
are nonzero; for the
fast/whistler case, only wavevectors at an angle $\le 45^\circ$ with
respect to the mean magnetic field $\V{B}_0$ are nonzero.  
The distribution of power is axisymmetric about $\V{B}_0$, with amplitudes
chosen to yield a one-dimensional magnetic energy spectrum
breaking from $k^{-5/3}$ to $k^{-2.8}$ at $k \rho_i=1$, consistent with
observations \citep{Alexandrova:2008,Podesta:2009,Sahraoui:2009}. 
%Although the solar wind is not axisymmetic, the axisymmetiric assumption 
%does not affect the findings herein.
Variation of these spectral indices within observational constraints
does not significantly impact our findings.

The synthetic time series are sampled on a $45^\circ$ trajectory with
respect to $\V{B}_0$.  For each turbulence model and wavevector range,
we choose five values of $\overline{V} \in [10.0,3.0,1.0,0.3,0.1]$
to model conditions in both the near-Earth solar wind and solar corona,
constructing an ensemble of $64$ time series with independent random
phases $\eta_{\V{k},m}$ for each case.  A Taylor hypothesis case is
also computed with $\omega_m(\V{k})=0$ in eq.~(\ref{eqn:Bt}).  The
magnetic energy frequency spectrum is calculated for each time series
and then ensemble averaged. These averages are shown in
Figs.~\ref{fig:overlay} and~\ref{fig:offset}, with the spectra of the
inertial and transition ranges overlaid.

%=====================================================================
\section{Results} 
\begin{figure}[t]
\hspace*{-0.0cm}
\includegraphics[scale=1.05,viewport=3 15 250 100, clip=true]
%{figs/overlay.eps}
{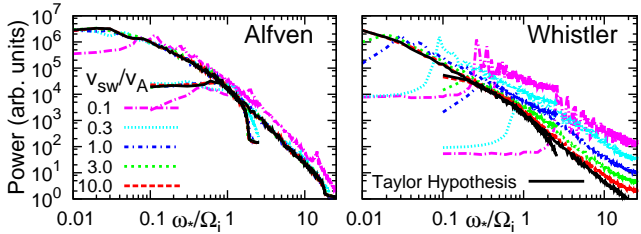}
\caption{ Averaged magnetic energy spectra from an ensemble of
  synthetic spacecraft data generated from critically balanced
  \Alfvenic (left) or  parallel fast/whistler (right) turbulence
  for 5 values of $\overline{V}$ (colors) and the Taylor
  hypothesis case (black). Inertial and transition range results are
  overlaid.  }
\label{fig:overlay}
\end{figure}

\begin{figure}[t]
%\hspace*{-0.5cm}
\hspace*{-0.25cm}
\includegraphics[scale=0.95,viewport=7 5 270 260, clip=true]
{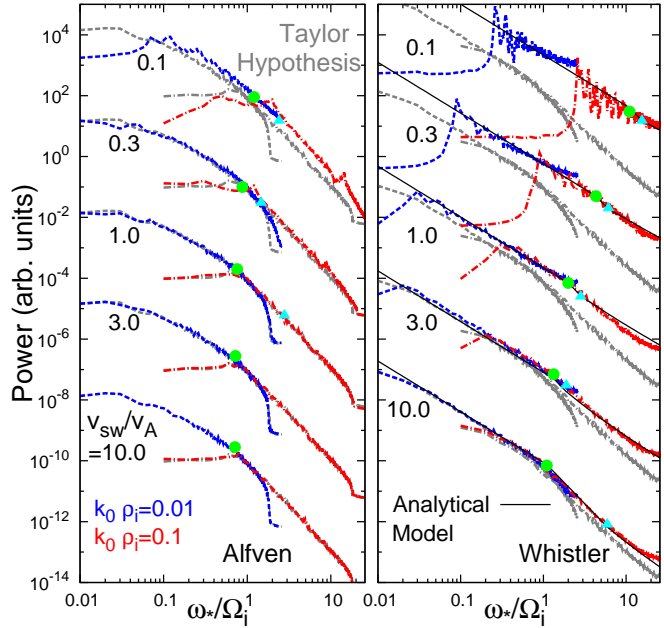}
%{figs/offset_pw_clean.eps}
%{figs/offset_pw.eps}
%{figs/offset_spec_prl.eps}
\caption{ The same spectra shown in Fig.~\ref{fig:overlay} offset
  %Averaged magnetic energy spectra from an ensemble of
  %synthetic spacecraft data generated from critically balanced
  %\Alfvenic (left) or  parallel fast/whistler (right) turbulence
  %offset
  vertically to highlight deviation of each case from the Taylor
  hypothesis case (grey). Inertial (blue dash) and transition (red
  dot-dash) ranges are shown. 
  In the right panel, the analytical model
  (black solid) is plotted, along with the break (green circle) and
  critical (cyan triangle) frequencies.
}
\label{fig:offset}
\end{figure}

In Fig.~\ref{fig:overlay}, the magnetic energy
spectra for the five $\overline{V}$ cases (colors) are compared to the
Taylor hypothesis case  (black).  
In Fig.~\ref{fig:offset}, each
$\overline{V}$ case is offset vertically for individual comparison to
the Taylor hypothesis case (grey, same vertical offsets), with
inertial range (blue) and transition range (red) results
distinguished. 

To facilitate the comparison between frequency spectra with
differing values of $\overline{V}$, we adopt the normalization
$\omega_*/\Omega_i\equiv (\omega_{sc}/\Omega_i)/\overline{V}$, which
transforms eq.~(\ref{eqn:map}) to 
\begin{equation}
\omega_*/\Omega_i = (\omega/\Omega_i)/\overline{V} + kd_i \cos \theta,
\label{eqn:map_norm}
\end{equation}
where $\V{k}\cdot
\V{v}_{sw} = kv_{sw} \cos\theta$. This transformation puts the
$\overline{V}$ dependence into the plasma-frame frequency term, so any
difference between the Taylor hypothesis case and a finite
$\overline{V}$ case is due to a violation of the Taylor hypothesis.

The primary qualitative results of this Letter are apparent in
Figs.~\ref{fig:overlay} and~\ref{fig:offset}. 
%Fig.~\ref{fig:offset}. 
For the critically
balanced \Alfvenic turbulence (left panels), there is no significant
violation of the Taylor hypothesis for flow velocity ratios
$\overline{V} \ge 0.3$. For the parallel fast/whistler turbulence, the
Taylor hypothesis is violated significantly for all values of
$\overline{V}$. These results confirm the analytical predictions in
Paper I.  The spacecraft-frame
frequency spectrum for the case of parallel fast/whistler turbulence
is modified qualitatively by two distinct effects\footnote{
Note that there
are no wavevectors satisfying the critical balance criteria 
at scales larger than $k_\perp \rho_i = 0.031$ due to our cubic domain restriction.
This lack of wavevector power for large scale \Alfven waves results in underfilled
lower frequency spectra, yielding a distinct curvature for
$\omega_*/\Omega_i \lesssim 0.3$ instead of the expected power law.
}.

The first effect is a shift of the entire spectrum from the Taylor
hypothesis case to higher frequency, seen for $\overline{V}\le 1$.
This effect is due to the slow flow of the solar wind with respect to
the \Alfven velocity, $v_{sw} \lesssim v_A$. The non-negligible
contribution from the plasma-frame frequency in eq.~(\ref{eqn:map}) leads
to this constant shift at all frequencies and is highlighted by a
shift to the right of the \emph{break frequency}, $\omega_{b*}$ (green
circle in Fig.~\ref{fig:offset}). In contrast, for  critically
balanced \Alfvenic turbulence, there is not a significant shift in
$\omega_{b*}$ for any value of $\overline{V}$.

The second effect is the flattening of the spectrum at high frequency,
most easily seen in the right panels of 
Figs.~\ref{fig:overlay} and~\ref{fig:offset} as
$\overline{V}$ decreases. This effect is due to the dispersive nature
of whistler waves at $k d_i \gtrsim 1$, leading to a more rapid than
linear increase of plasma-frame frequency with increasing
wavenumber. Since the spatial advection term scales linearly with
wavenumber, even for rapid solar wind flow, eventually the
plasma-frame frequency term will become non-negligible due to its more
rapid increase with wavenumber. This dispersive increase of the wave
frequency flattens the frequency spectrum for frequencies higher than
 a \emph{critical frequency}, $\omega_{c*}$ (cyan triangle in Fig.~\ref{fig:offset}),
although this transition is often gradual.

Several other minor features are apparent in Figs.~\ref{fig:overlay}
and~\ref{fig:offset}. First, the pronounced peaks in the low
$\overline{V}$ cases are an artifact caused by the discrete nature of
our wavevector grid at the largest scales.  Second, the contribution
from the plasma-frame frequency shifts power to higher frequencies,
causing a small but noticeable aliasing at the high-frequency end of
the spectra as more power is shifted above the Nyquist frequency of
the synthetic time series.  Finally, small bumps in the spectra for the
\Alfvenic cases with $\overline{V} \le 0.3$ at $\omega_*/\Omega_i \sim
10$ are due to a mode transition from kinetic \Alfven waves to ion
Bernstein waves in our turbulent power distribution at $k_\perp \rho_i
\simeq 7$.

\section{Analytical Model}
To illuminate the effect of the violation
of the Taylor hypothesis, we construct a simple analytical model that
reproduces the two primary qualitative effects in the slow flow and
dispersive regimes. Here we present only the model for the parallel
fast/whistler turbulence. 

For a given piecewise-continuous, magnetic energy wavenumber spectrum
$E(k)=(E_0/k_0) (kd_i)^{-5/3}$ at $kd_i <1$ and $E(k)=(E_0/k_0)
(kd_i)^{-2.8}$ at $kd_i \ge 1$, the problem boils down to finding a
mapping $k(\omega_{sc})$ so that we may
determine the frequency spectrum $E(\omega_{sc})=E[k(\omega_{sc})]$.
Here $k_0$ is the outer scale wavenumber and  $E_0$ is a constant to adjust the total turbulent magnetic energy. 
%over $k_0\le k \le \infty$ is $E=E_0[2(k_0d_i)^{-5/3}/3-38(k_0d_i)^{-1}/30]$. 
The fast/whistler wave dispersion relation is expressed by
$\omega=kv_A\sqrt{1+(k_\parallel di)^2}$ in Paper I.

The key simplification needed to obtain an analytical solution is to
find the mapping $k(\omega_{sc})$ along a particular 1D path through
3D wavevector space.  We choose the path $k_\perp =k_\parallel$, and we
sample at velocity $\V{v}_{sw}$ along this path at a $45^\circ$ angle
with respect to $\V{B}_0$; in this case, $\theta=0$, so $\V{k}\cdot
\V{v}_{sw}=k v_{sw}$. Although this choice may seem to limit the
generality of the solution, the steeply dropping energy spectrum is
dominated by the largest frequency associated with a particular
wavevector amplitude $k$; the case $\theta=0$ gives the maximum
frequency from the advection term for a given $k$, and therefore this
is the dominant contribution. Along this path, eq.~(\ref{eqn:map_norm})
becomes
\begin{equation}
\omega_*/\Omega_i = kd_i\left[1
  +\sqrt{1+(1/2)(kd_i)^2}/\overline{V}\right].
\label{eq:map_model}
\end{equation}

We convert this function into the piecewise function
$\omega_*/\Omega_i = kd_i(1 +1/\overline{V})$ for $kd_i<\sqrt{2}$ and
$\omega_*/\Omega_i = kd_i[1 +kd_i/(\overline{V}\sqrt{2})]$ for $kd_i
\ge \sqrt{2}$. This piecewise function may be easily inverted to yield,
\begin{equation}
kd_i=\left\{ \begin{array}{lc}
\frac{\omega_*}{\Omega_i}\frac{\overline{V}}{1+\overline{V}}, & \frac{\omega_*}{\Omega_i} < \sqrt{2}(1+\frac{1}{\overline{V}})\\
-\frac{\overline{V}}{\sqrt{2}} + \sqrt{\frac{\overline{V}^2}{2}+\sqrt{2}\frac{\omega_*}{\Omega_i} \overline{V}}, & \frac{\omega_*}{\Omega_i} \ge \sqrt{2}(1+\frac{1}{\overline{V}}).
\end{array}
\right.
\end{equation}

Using this function for $k(\omega_*)$, we can immediately plot the
frequency spectrum $E(\omega_*)$, shown as the solid black lines in
the right panel of Fig.~\ref{fig:offset}. This simple analytical
model agrees well with the frequency spectra generated by the
synthetic spacecraft data method at all values of $\overline{V}$. In
addition, the model may be used to obtain analytical estimates for the
break frequency $\omega_{b*}$ and the critical frequency $\omega_{c*}$. The
break frequency for our model occurs at $k d_i=1$, so we obtain
$\omega_{b*}/\Omega_i= 1+1/\overline{V}$, given by the green circles
in Fig.~\ref{fig:offset}.  The critical frequency, where dispersive
effects become significant, requires both the waves to be dispersive
and the plasma-frame frequency term to be significant (taken to be
$|\omega| \ge |\V{k}\cdot \V{v}_{sw}|/3$). This concurrence occurs 
at $k d_i = \sqrt{2} \max[1,((\overline{V}/3)^2-1)^{1/2}]$ and leads to a
prediction for the critical frequency of $\omega_{c*}/\Omega_i=
\sqrt{2}(1+1/\overline{V})$ for $\overline{V}\le 3 \sqrt{2}$ and
$\omega_{c*}/\Omega_i= \sqrt{2}[((\overline{V}/3)^2-1)^{1/2} +
  ((\overline{V}/3)^2-1)/\overline{V}]$ for $\overline{V}> 3
\sqrt{2}$, indicated by cyan triangles in Fig.~\ref{fig:offset}.

\section{Discussion} We can use this analytical model to predict
quantitatively the effect of the violation of the Taylor hypothesis on
the magnetic energy frequency spectrum, as depicted in
Fig.~\ref{fig:line}. The violation in the slow flow regime is easily
calculated in the limit $kd_i \ll 1$, simplifying
eq.~(\ref{eq:map_model}) to $\omega_*/\Omega_i \simeq kd_i(1
+1/\overline{V})$.  Since the scaling of $\omega_*$ is the same as
$k$, the spectrum will have the same scaling but will be shifted to
higher frequency by a factor $(1 +1/\overline{V})$, as depicted by the
blue arrows in Fig.~\ref{fig:line}. To highlight this effect for
non-dispersive waves, we apply this result to an artificial
fast/whistler dispersion relation that has no dispersion,
$\omega/\Omega_i=k d_i$ (blue dot-dash).

To calculate the violation of the Taylor hypothesis in the dispersive
regime, we simplify eq.~(\ref{eq:map_model}) in
the limit $kd_i \gg 1$ to $\omega_*/\Omega_i \simeq kd_i[1
  +kd_i/(\overline{V}\sqrt{2})]$. When $kd_i \gg
\overline{V}\sqrt{2}$, the plasma-frame frequency term dominates due
to the more rapid than linear increase of whistler wave frequency with
wavenumber, $\omega/\Omega_i=k d_i\sqrt{1+(k d_i)^2/2}$, giving
$\omega_*/\Omega_i \simeq (kd_i)^2/(\overline{V}\sqrt{2})$. Therefore,
we obtain a mapping $k \propto \omega_*^{1/2}$, leading to a
flattening of the frequency spectrum to $E(k) \propto k^{-2.8} \propto
\omega_*^{-1.4}$, as indicated by the red arrows in
Fig.~\ref{fig:line}. Note, however, that the onset of this
flattening can be gradual, only reaching $E(k) \propto
\omega_{sc}^{-1.4}$ at $kd_i \gg \overline{V}\sqrt{2}$, as seen in
Fig.~\ref{fig:line}.

\begin{figure}[t]
\hspace*{-0.5cm}
\includegraphics[scale=0.95,viewport=0 15 275 115, clip=true]
%{figs/compensated_new.eps}
%{figs/compensated_prl.eps}
{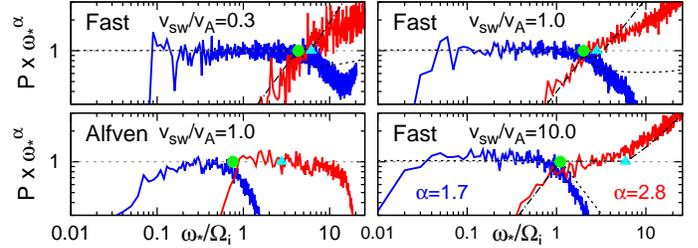}
\caption{ Compensated magnetic energy spectra from four synthetic data
  sets generated for \Alfvenic turbulence with $\overline{V}=1.0$
  (lower left) and fast/whistler turbulence with
  $\overline{V}=10.0,\ 1.0,\ 0.3$ (counter-clockwise from lower
  right). Spectra from the inertial and transition ranges are
  compensated by $\omega_*^{1.7}$ (blue) and $\omega_*^{2.8}$
  (red). Also plotted are the analytical model compensated by
  $\omega_*^{1.7}$ (black dash) and  $\omega_*^{2.8}$ (black dot-dash), as well as 
  $\omega_{b*}$ (green circle) and $\omega_{c*}$ (cyan triangle).  }
\label{fig:comp}
\end{figure}

To highlight further the features of the model and to compare to the
synthetic spacecraft data results, we plot compensated spectra in
Fig.~\ref{fig:comp}. The magnetic energy frequency spectra produced
from the inertial range (blue) and transition range (red) are
compensated by $\omega^{1.7}_*$ and $\omega^{2.8}_*$, respectively.
For three parallel fast/whistler cases with
$\overline{V}=10.0,1.0,0.3$, (counter-clockwise from lower right) the
compensated energy spectra are flat up to $\omega_{c*}$,
and steepen at higher frequencies.
The values for $\omega_{b*}$ and $\omega_{c*}$
calculated from the model correspond well with the breaks seen in the
compensated synthetic energy spectra.
The strong correspondence between
the one-dimensional analytic model and the three-dimensional
synthetic spacecraft results serve as \emph{a posteriori} support for 
the approximations used in the analytic model.

Note that synthetic time series with $\beta_i \ll 1$ (not shown
here), relevant to the near-Sun environment, have been 
found to have qualitatively similar spectral features
to those described here.

\section{Conclusions} In this Letter, we determine the qualitative
and quantitative effects on the measured magnetic energy frequency
spectrum in the solar wind due to the violation of the Taylor
hypothesis. For upcoming spacecraft missions, such as \emph{Solar
  Probe Plus}, we find the Taylor hypothesis may be violated in two
regimes: the slow flow and dispersive regimes. In the slow flow regime, a
significant plasma-frame frequency contribution to the
spacecraft-frame frequency leads to a shift of the frequency spectrum
to higher frequency by a factor $1 + v_A/v_{sw}$ relative to a Taylor
hypothesis case where $\omega_{sc}=\V{k}\cdot \V{v}_{sw}$ but no
change in the scaling of the spectrum. Since the underlying wavevector
spectrum cannot be determined by a single spacecraft, this effect is
undetectable. In the dispersive regime, the dispersive increase of
wave frequency with wavenumber, $\omega \propto k^2$, can lead to a
flattening of the typical dissipation range $k^{-2.8}$ wavenumber
spectrum to an $\omega^{-1.4}$ frequency spectrum.  We confirm earlier
predictions from Paper I that critically balanced \Alfvenic
turbulence will not, but parallel fast/whistler turbulence will,
significantly violate the Taylor hypothesis, especially near the
\Alfven critical point where $v_{sw} \sim v_A$. Thus, a flattening of
the frequency spectrum in the dissipation range is the predicted
observational signature for fast/whistler turbulence. For \emph{Solar
  Probe Plus} measurements, the shifting of power to higher
frequencies may also threaten to cause significant aliasing of
measured signals, even at the high sampling rate of the instruments.

This work was supported by NSF CAREER AGS-1054061 and NASA NNX10AC91G.

%-=-=-=-=-=-=-=-=-=-=-=-=-=-=-=                                                 
\bibliographystyle{apj}

%\bibliography{../../bibs/abbrev,../../bibs/HeliophysicsTheory,../../bibs/HeliophysicsObservation,../../bibs/Simulations,../../bibs/KineticTheory,../../bibs/Turbulence,../../bibs/preparation.bib}

%\bibliography{../../bibs/abbrev,../../bibs/HeliophysicsTheory,../../bibs/HeliophysicsObservation,../../bibs/Simulations,../../bibs/KineticTheory,../../bibs/Turbulence,../../bibs/preparation.bib}
%-=-=-=-=-=-=-=-=-=-=-=-=-=-=-=    

\end{document}